\documentclass[aps,reprint,pra,superscriptaddress,nofootinbib]{revtex4-2}
\usepackage{amsmath,amsfonts,amssymb}
\usepackage{mathtools}
\usepackage[pdftex]{graphicx}
\usepackage[dvipsnames, svgnames, table]{xcolor}
\usepackage[colorlinks=true, citecolor=Blue,linkcolor=BrickRed, urlcolor=ForestGreen]{hyperref}
\usepackage{txfonts}
\usepackage{mathrsfs}
\usepackage{bm}
\usepackage{multirow}
\usepackage{braket}
\newcommand{\be}{\begin{equation}}
\newcommand{\ee}{\end{equation}}
\newcommand{\ben}{\begin{eqnarray}}
\newcommand{\een}{\end{eqnarray}}
\newcommand{\bes}{\begin{subequations}}
\newcommand{\ees}{\end{subequations}}
\newcommand{\bF}{\begin{figure}}
\newcommand{\eF}{\end{figure}}

\begin{document}

\title{LEVITAS: Levitodynamics for Accurate Individual Particle Sensing in Space}

\author{Rafal Gajewski}
\affiliation{Department of Physics, University of Warwick, Coventry CV4 7AL, United Kingdom}

\author{Ravindra T Desai}
\email{ravindra.desai@warwick.ac.uk}
\affiliation{Department of Physics, University of Warwick, Coventry CV4 7AL, United Kingdom}

\author{James Bateman}
\email{j.e.bateman@swansea.ac.uk}
\affiliation{Department of Physics, Swansea University, Swansea, SA2 8PP, United Kingdom}

\author{Bengt Eliasson}
\affiliation{SUPA Department of Physics, University of Strathclyde, Glasgow, G4 0NG, United Kingdom}

\author{Daniel K L Oi}
\email{daniel.oi@strath.ac.uk}
\affiliation{SUPA Department of Physics, University of Strathclyde, Glasgow, G4 0NG, United Kingdom}

\author{Animesh Datta}
\email{animesh.datta@warwick.ac.uk}
\affiliation{Department of Physics, University of Warwick, Coventry CV4 7AL, United Kingdom}

\date{\today}

\begin{abstract}

Accurately observing the rarefied media of the upper atmosphere, exosphere, and planetary and solar system environments and beyond requires highly sensitive metrological techniques.
We present the operating concept and architecture of an in-situ sensing solution based on the dynamics of a levitated nanoparticle (levitodynamics).
It can detect and measure impacts of individual particles in rarefied media. 
Dubbed `LEVITAS', our sensor consists of a dispenser of dielectric nanoparticles and optical trapping of a single nanoparticle in the focus of a laser beam. The trapped nanoparticle constitutes a harmonic oscillator at frequencies in the kilohertz range whose position can be tracked at the standard quantum limit by interferometric detection of the laser photons it scatters.
Here, we simulate microcanonical impacts on the nanoparticle and show that the density, velocity, temperature, and composition of the surrounding medium can be estimated accurately.
We illustrate the performance of LEVITAS in circumstances ranging from low Earth orbit out to exospheric distances, across which individual impacts can be detected at favourable rates.
Furthermore, LEVITAS may be employed to accurately measure highly rarefied neutral distributions within vastly different areas of momentum space. This we demonstrate by simulating the measurement of high-velocity neutral gas particles from the interstellar medium penetrating the heliosphere and flowing through our solar system.

\end{abstract}

\maketitle

\section{Introduction} 
\label{sec:intro}

The in-situ detection of sparse distributions of atoms, molecules and dust in space has a wide range of applications.   
Exploratory probes sent to the distant reaches of the solar system and beyond have revealed the fundamental constituents, dynamics, and evolution of planetary atmospheres and exospheres, comets, interplanetary material, as well as in the very local interstellar medium \citep{Rhea,ISM,Mars,Venus,Moon,IPDust,Saturn,Comet,Jupiter,Mercury}.
Space-borne measurements of our own upper atmospheric and exospheric constituents have enabled us to understand the complex interaction of the Sun and magnetosphere with our atmosphere and climate. They can also directly support space weather forecasting by understanding how resultant atmospheric variabilities modulate satellite orbits to deliver accurate collision risk assessments and warnings to satellite operators \citep{Berger2020,Elvidge2019,Lin2022,Milan2017,Picone2002}.

These diverse measurements are conducted using a wide variety of technological solutions. For example, the majority of in-situ measurements of neutral populations in the Earth's upper atmosphere derive from accelerometers \citep{Li2022,Mehta2017,Siemens2016,Woeske2024}.
These cannot resolve the composition or higher moments of the distribution functions. 
Mass spectrometers have provided additional constraints up to several hundred kilometers \citep{Banks1981,Hanson1975,Philbrick1974}, and also of neutral populations on other planets \citep{Mars,Waite2005}. These typically operate through ionising incoming neutrals and detecting the ionised products. However, significant calibration is required, particularly in low-density regimes \citep{DeKeyser2019_CalibrationDFMS,Mogul2023_LNMSReprocessing, Perry2015,Teolis2015}. 
For altitudes greater than several hundred kilometers, the measurement regime switches to that of Lyman-alpha absorption which can provide broad large-scale integrated line-of-sight densities \citep{Baliukin2019,Schuhle2011}.
Particles with higher momenta, such as interstellar neutrals ~\citep{fuselier2009ibexlo},  energetic neutral atoms~\citep{Williams1992,Gruntman1997,Brandt2021}, and nanometre dust \citep{OBrien2014, Xie2018} are similarly detected via indirect means, typically through surface-interaction techniques in which incident particles eject electrons or ions from a target and the resulting electrical signals are used to derive information on the incident particles.

The ability to directly and accurately detect and measure \emph{individual} particles in space, across a wide range of environments in terms of densities and momenta with minimal reliance on in-flight calibration, would greatly benefit future space missions and open new avenues for scientific enquiry. 

\begin{figure}
    \includegraphics[width=0.475\textwidth]{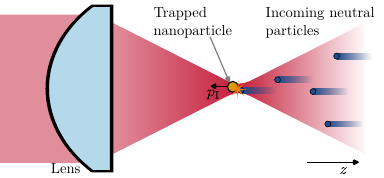}
    \caption{Schematic demonstrating the operation of LEVITAS in one dimension; a dielectric nanoparticle is trapped at the focus of a laser beam, and exposed to a rarefied medium outside the spacecraft. Collisions with individual neutral species therein lead to nanoparticle recoil with momentum $p_\text{I}$ at time $t_\text{I}$. The objective of LEVITAS is to provide their estimates $p'_\text{I}$ and $t'_\text{I}$ respectively.
    These can be used to estimate the density, velocity, temperature, and composition of the target medium or enable high fidelity reconstructions of the underlying distribution function itself. See Fig.~\ref{fig:design} for the full sensor head.}
    \label{fig:levitas_demo}
\end{figure}

Thus motivated, we present a new sensor for detecting individual particles in space using the dynamics of a levitated nanoparticle (levitodynamics). See Fig.~\ref{fig:levitas_demo} for a schematic. We dub our sensor `LEVITAS', and provide a theoretical analysis of its application to detecting neutral particles.
Charged particles are prevented from hitting the nanoparticle using electrostatic deflector plates, so that plasma effects on the nanoparticle \cite{Barnes1992,Shukla2009} can be neglected.
LEVITAS is a quantum sensing approach to rarefied environment characterization. Recent cold atom vacuum standards (CAVS)~\citep{frieling2024crosscalibration, booth2024revising,eckel2025effect} have demonstrated that quantum systems can achieve precision pressure measurements: laser-cooled atoms, typically $^7$Li or $^{87}$Rb are held in shallow traps, and the rate at which atoms are ejected provides a measure of ambient pressure, achieving percent-level accuracy for known gas species.

LEVITAS extends this quantum sensing paradigm by 
optically tracking a trapped nanoparticle's position in real time with a precision limited by quantum mechanics. Doing so in the aftermath of collision-induced momentum impacts -- on a typically nano- to micrometre-sized dielectric particle, allows LEVITAS to quantify the recoil momentum $p_\text{I}$ of each impact on the nanoparticle. 
This, combined with the relative speed between a spacecraft and the species in the medium, allows direct in-situ detection of individual microcanonical impacts. These can then be used to infer the density, velocity, temperature, and composition of the neutral gas species. LEVITAS can thus provide insight and information about phenomena that were hitherto unavailable.

Rarefied space environments span a wide range: from low Earth orbit (LEO), where oxygen atoms encounter the spacecraft at $\sim 7.5$~km/s imparting momenta\footnote{We present momenta in units of u km/s, where u denotes one atomic mass unit of $1.66 \times 10^{-27}$ kg. Thus, $1$ u km/s $= 1.66 \times 10^{-24}$ kg m/s  = $3.11$ keV/c.}  of $\sim120$~u~km/s, to the local interstellar medium penetrating and flowing through our solar system, where hydrogen molecules at $\sim25$~km/s impart momenta of $\sim50$~u~km/s. To put these into perspective, levitated micron-sized nanoparticles in the laboratory have already detected individual nuclear $\alpha$-decays with momenta of $\sim 60$ u km/s~\citep{Moore2024}, while 
nanoparticles of radius $\sim$ 85 nm have detected momenta of $\sim$ 40 u km/s~\citep{tseng2025search}, indicating the experimental feasibility of LEVITAS.

This paper is organized as follows. In Sec.~\ref{sec:inst}, we describe the operating principle of LEVITAS and the theoretical framework for detecting individual, microcanonical momentum impacts on the nanoparticle. In Sec.~\ref{sec:gas_dynamics}, we develop a Bayesian inference framework for extracting thermodynamic properties such as density, temperature, velocity, and composition from the measured recoil momenta. In Sec.~\ref{sec:results}, we demonstrate the performance of LEVITAS through simulated measurements of low Earth orbit environments at altitudes of 600 km and 1000 km, and interstellar neutral flows at the second Earth-Sun Lagrange point L$_2$, about $1.5\times10^6$~km away from the Earth on the opposite side from the Sun. In Sec.~\ref{sec:design}, we present an architecture of the LEVITAS sensor head, including the nanoparticle delivery and optical trapping systems. In Sec.~\ref{sec:extensions}, we discuss extensions to LEVITAS such as a pressure-mode operation at higher densities, multiplexing strategies for enhanced throughput in extremely rarefied media, different nanoparticle geometries for improved directional sensitivity, as well as quantum techniques to detect extremely low momentum impacts. We conclude in Sec.~\ref{sec:disc} by noting some limitations and future prospects of LEVITAS.

\section{Detecting individual impacts} 
\label{sec:inst}

LEVITAS is based on recent advances in levitodynamics~\citep{gonzalezballestero2021levitodynamics}, which studies the levitation and control of microscopic objects in vacuum. Experiments on optically levitated dielectric nanoparticles have been developed from initial demonstrations of feedback cooling~\citep{gieseler2012subkelvin} to quantum-limited sensors capable of precise force measurements at the $\sim 10^{-21}$~N~\citep{ranjit2016zeptonewton} and $\sim 10^{-24}$~N~\citep{liang2023yoctonewton} scale. Experiments with $\sim 100$~nm silica nanoparticles have demonstrated ground state cooling in both optical cavities~\citep{delic2020cooling} and free space configurations~\citep{Magrini2021, tebbenjohanns2021quantum}, achieving quantum-limited control of the centre-of-mass motion.
Real-time state estimation and control using digital feedback systems~\citep{Setter_2018} enables continuous monitoring at this limit.

The progression from basic optomechanical control towards momentum detection was studied theoretically for collision-resolved pressure sensing~\citep{barker2024collisionresolved}, which details how quantum-limited nanomechanical devices could detect ultra-low gas pressures by counting individual particle collisions. Detecting the recoil from nuclear $\alpha$-decay~\citep{Moore2024} with micron-sized particles demonstrates this principle albeit with significantly larger momenta.
More recent work with nanoparticles under high vacuum approaches the standard quantum limit~\citep{tseng2025search}. LEVITAS advances these quantum-limited measurement techniques for in-situ  sensing of individual particles in space.
In so doing, it provides uniquely new capabilities, including neutral mass spectroscopy.

\subsection{Model of the nanoparticle dynamics}
Assuming a LEVITAS instrument installed on a spacecraft, the sensor head consists of a spherical nanoparticle of mass $M$ trapped at the focus of a laser beam aligned along the direction of motion of the spacecraft, taken to be along the $z$ axis; see Fig.~\ref{fig:levitas_demo}. In this work, we restrict ourselves to a one-dimensional setting to convey the operating principle of LEVITAS most directly. The dynamics of the trapped nanoparticle is modelled as a stochastically driven, damped harmonic oscillator. Each component of the nanoparticle displacement $\mathbf{r}=(x,y,z)$ from the trap centre obeys a decoupled Langevin equation.
The motion along the beam axis ($z$) is governed by
\begin{equation}
    \ddot{z}(t) + \gamma\dot{z}(t) + \Omega^2 z(t) = \frac{1}{M}F(t)\,,
    \label{eom}
\end{equation}
where $\gamma$ represents mechanical damping, $\Omega$ is the trap frequency along the beam axis, and $F(t) = \eta(t) + F_\text{I}(t)$ corresponds to the total external force, which includes random force fluctuations $\eta(t)$ and any force impulses due to individual impacts $F_\text{I}(t)$ that we seek to detect. The impact of individual neutral species on the nanoparticle is given by $F_\text{I}(t)=p_\text{I}\delta(t-t_\text{I})$ with momentum $p_\text{I}$ at time $t_\text{I}$, where $\delta(\cdot)$ is the Dirac delta function. The objective of LEVITAS is to provide their estimates $p'_\text{I}$ and $t'_\text{I}.$

We model the random force fluctuations as white noise having the autocorrelation function
\begin{equation}
    \langle\eta(t)\eta(t')\rangle = 2\pi S_\text{FF}\delta(t-t')
\end{equation}
with a power spectral density $S_\text{FF}$. 
We assume that the nanoparticle is trapped in a highly rarefied environment, such that the force fluctuations along the beam axis are dominated by laser recoil fluctuations with spectral density~ \citep{Tebbenjohanns2019}
\begin{equation}
S_\text{FF} = \left(\frac{2}{5} + A^2\right)\frac{\hbar\omega_0}{2\pi c^2}P_s
\label{eq:force-spectral-density}
\end{equation}
for laser frequency $\omega_0$, scattered power $P_s$, and a geometrical factor $A$ corresponding to the degree of beam focusing. To access the regime where photon recoil noise dominates, the effect of laser intensity noise must be reduced; intensity noise affects nanoparticle motion as a \emph{parametric} noise source \citep{savard1997heating} and hence can be rendered negligible with modest feedback cooling of the motion \citep{Jain_2016}.

We consider a spherical nanoparticle made of silica. The nanoparticle and trap parameters we use are given in Table~\ref{table:sim_params}.
For the latter, laser intensity noise becomes negligible when the centre-of-mass (c.m.) motion is cooled to $T_\text{CM} \sim 1\text{K}$ (see Appendix~\ref{sec:Tcm_reqs}), which for our parameters corresponds to a damping rate of $\gamma=2\pi\times 1.7\text{ mHz}$ due to feedback cooling \citep{Jain_2016}. The effective damping rate due to this cooling process dominates over all other processes.

\begin{table} 
\begin{center}
\begin{tabular}{lll}
\textbf{Parameter} & \textbf{Description} & \textbf{Value} \\ \hline\hline
$R$             & Nanoparticle radius           & 50 nm                           \\
$\rho$          & Silica density                & 2.3 g/cm$^3$                    \\
$M$             & Nanoparticle mass             & 7.3 $\times 10^8~\text{u}$      \\
$T_\text{CM}$   & Centre-of-mass temperature    & 1 K                             \\
$\Omega$        & Angular trap frequency        & $2\pi~\times~$12 kHz              \\
$\gamma$        & Angular damping rate          & $2\pi~\times~$1.7 mHz              \\
$\lambda$       & Laser wavelength              & 1550 nm                         \\
$P$             & Laser power                   & 500 mW                          \\
$P_s$           & Scattered power               & $\sim$5 nW                      \\
NA              & Numerical aperture            & 0.4                             \\
$B$             & Sampling rate                 & 1 MS/s                          \\
\hline
$v_s$           & Spacecraft speed              &                                 \\
$ $             & LEO @ 600 km altitude         & 7.5 km/s                        \\
$ $             & LEO @ 1000 km altitude        & 7.3 km/s                        \\
$ $             & L$_2$ (Interstellar medium)   & 26 km/s                         \\
\hline\hline
\end{tabular}
\end{center}
\caption{Typical nanoparticle and optical trap parameters in levitodynamics experiments with LEVITAS. Numerical aperture (NA) refers to the objective used for trapping. Speeds of spacecraft carrying LEVITAS are used in Sec.~\ref{sec:results}.}
\label{table:sim_params}
\end{table}

Our estimation of the
momentum $p_\text{I}$ and time $t_\text{I}$ begins by tracking the motion of the nanoparticle. We do so  with a linear position readout provided by the interferometric measurement of the back-scattered light. We assume that the photodetector is limited by shot noise and the numerical aperture of the trapping objective~\citep[Appendix D.2]{Tebbenjohanns2019}.

We use a Kalman filter to track the state of the nanoparticle (position $z$ and velocity $v_z$) and identify the impacts. The state vector also  includes the impulse force $F_\text{I}$; 
see Appendix \ref{sec:kalman_filter} for details.
The estimate of the momentum $p_\text{I}$ is computed with a moving average $\tilde{p}_\text{I}=\int_\Delta F_\text{I}' \text{d}t$ over a small time window $\Delta$, where $F_\text{I}'$ denotes the Kalman estimate of the impact force $F_\text{I}$. The occurrence of impacts is indicated by the spikes in the resulting time trace. These are used as initial guesses for Bayesian estimation (cf.~Appendix \ref{sec:bayesian}) of the recoil momentum and the corresponding confidence intervals. Specifically, the position signal data in a small window around each flagged time is used to maximise the logarithm of the posterior probability distribution derived from Eq.~\eqref{dynamics_model_bayesian},
marginalised over the nuisance parameters.

\subsection{Simulating detection of individual impacts} \label{sec:simulation}

We start by calculating the nanoparticle's trajectory by solving Eq.~\eqref{eom} numerically using the Euler--Maruyama method to obtain particle displacement along the beam ($z$) axis. We model impacts as discrete changes in the nanoparticle's velocity at particular impact times. To simulate the corresponding experimental signal, we sample the trajectory at the chosen measurement sampling rate, with additive white Gaussian measurement noise, assuming that the interferometric readout is calibrated to provide a direct measurement of the position.

\subsection{Recoil momentum inference} \label{sec:tracking}
\begin{figure}
    \centering
    \includegraphics[width=1.0\linewidth]{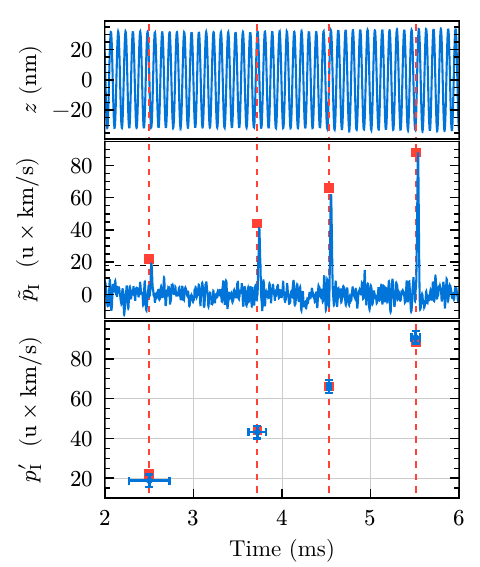}
    \caption{Recoil momentum estimation from a trajectory subject to four individual impacts, simulated for the parameters in Table~\ref{table:sim_params}. Top: A record of the nanoparticle's dispacement. Middle: a Kalman filter is used to obtain a moving average (over a quarter the trap cycle) estimate of the recoil momentum $\tilde{p}_\text{I}$, with impact times flagged by peaks above the threshold level $p_{\text{th}} =$ 18 u km/s denoted by the black dashed line. Bottom: flagged impact times are used as initial guesses for a Bayesian inference calculation, used to compute final estimates of $t_\text{I}'$ and $p_\text{I}'$ and the corresponding errors. Red squares and dashed lines mark the momenta and times of the true impacts respectively. The horizontal errors in the impact times are magnified to represent $100\sigma$.}
    \label{fig:kick-demo}
\end{figure}

We now present the procedure to obtain estimates $p'_\text{I}$ and $t'_\text{I}$ of  $p_\text{I}$ and $t_\text{I}$ respectively, as shown in Fig.~\ref{fig:kick-demo}.
Note that the peaks in the moving average (middle panel) of the momentum $\tilde{p}_\text{I}$ obtained from the Kalman filter appear with a slight delay behind the true impact time. 
We calibrate this delay, so that we can correct it before the Bayesian inference step, using known impact times and observing the delay.
This calibration procedure can be experimentally replicated using electrical impulses \citep{Moore2024}.
For the simulation parameters in Table~\ref{table:sim_params}, we find that the peaks lag behind the true impact times by about 5 \textmu s.

\begin{figure}
    \centering
    \includegraphics[width=1.0\linewidth]{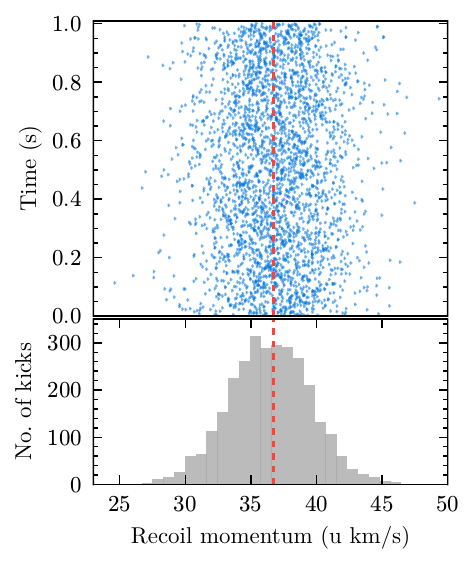}
    \caption{Detection of individual impacts applied uniformly in time at a rate 3000/s, and identical momentum of $p_\text{I}=$ 36.73~u~km/s (denoted with the red dashed line). The error bars on the points in the top plot are hidden for clarity. The histogram contains 2996 events with sample mean $36.84\pm3.15$ u km/s and mean recoil estimated error of $\sigma_\text{det}=3.15\pm0.03$~u~km/s. The momentum threshold was set to $p_{\text{th}} =$ 18 u km/s as in Fig.~\ref{fig:kick-demo}.}
    \label{fig:v50_histogram}
\end{figure}

To assess the momentum resolution of the recoil measurement, we use the simulation procedure described above with the impacts applied at a rate of 3000/s and a recoil momentum of $p_\text{I}=$ 36.73 u km/s. For comparison, a helium atom moving at a relative speed of 7.8 km/s imparts a recoil of 31.2 u km/s. In this Section, the number of collisions occurring in a given time interval is taken to be a constant, but when inferring the thermodynamic properties of the atmospheric gas in Sec.~\ref{sec:results}, the number of collisions occurring in a given time interval is modelled using a Poisson distribution.

The results for around one second of simulated time are shown in Fig.~\ref{fig:v50_histogram}. The detected impacts appear normally distributed around the true impact magnitude, with about 0.3\% accuracy and 8.6\% precision. Each recoil event was recorded with an error of $\sigma_\text{det}=3.15\pm0.03$ u km/s, consistent with the standard deviation of the whole sample. 

We used a simple approach where only impacts above the momentum threshold of $p_{\text{th}} =$ 18 u km/s
were accepted. Then 2996 impacts out of 2997 were detected, with no false detections. Testing over longer simulation runs with no impacts applied yielded a rate of $\sim 1$/s false detections at this threshold level. The appropriate threshold level depends on the expected impact rate and therefore the density of the medium. In extremely rarefied media, where the expected impact rate based on the nanoparticle is $\lesssim$ 1/s, it is imperative that the expected rate of false detections remains negligible, as discussed in Sec.~\ref{sec:disc}. Additional demands on the signal shape using, for example, the matched filter approach \citep{Moore2024}, could allow detection of lower magnitude impacts without a detrimental increase in the rate of false detections.

\section{Inferring thermodynamic properties}
\label{sec:gas_dynamics}
 
The detection scheme in Sec.~\ref{sec:inst}
allows LEVITAS to build up a distribution of the recoil momenta which can then be used to infer the thermodynamic quantities of the medium.

We consider a gas that moves with a relative speed of $u= v_s+w$ along the direction of motion of the spacecraft, where $v_s > 0$ is the spacecraft speed and $w$ is the wind speed of the gas, where
$w>0$ represents flow towards the spacecraft. The speed $v$ of each species of mass $m_i$ in the medium is assumed to follow a shifted Maxwell-Boltzmann distribution in one dimension $\mathcal{N}(v-u,\sigma^\text{MB}_i)$ at temperature $T$,  $\sigma^\text{MB}_i=\sqrt{k_B T/m_i.}$ Here,  $i=0,\cdots,S-1$ denotes $S$ different species with densities $n_i,$ with a total number density $n= \sum_{i=0}^{S-1} n_i,$ and
$\mathcal{N}(v-u,\sigma) =(1/\sqrt{2\pi\sigma^2}) \exp(-(v-u)^2/2\sigma^2)$. See Sec.~\ref{sec:disc} for extensions to other known or unknown distributions in one or more dimensions.

The trapped nanoparticle facing the gas distribution receives impacts at a differential rate $r$ determined by its cross section
\begin{equation}
    \frac{\text{d} r(v)}{\text{d}v} = \pi R^2 v\sum_{i=0}^{S-1} n_i \mathcal{N}(v-u,\sigma^\text{MB}_i)\Theta(v)\,,
    \label{eq:differential_rate}
\end{equation}
where 
$\Theta$ represents the Heaviside step function. Upon integrating Eq.~\eqref{eq:differential_rate} with $u\gg\sigma_i$, which we assume to be the case here for the relevant temperatures and spacecraft speeds\footnote{The full solution has a pre-factor $f(u/\sigma_i^\text{MB})$ where $f(x)=\frac{1}{x\sqrt{2\pi}}\exp(-x^2/2)+\frac{1}{2}[1+\text{erf}(x/\sqrt{2})]$. For $u=7.8$ km/s and for highest thermal spread considered here, hydrogen at 5000~K, $\sigma^\text{MB} = 6.44$ km/s, the rate without the pre-factor differs by less than 5\%.},
the total rate is $r\approx n\pi R^2u$.
We thus have
\begin{equation}
    \text{prob}(v|\bm{\theta}) = \frac{1}{r}\frac{\text{d} r(v)}{\text{d}v} = \frac{v}{u}\sum_{i=0}^{S-1} c_i \mathcal{N}(v-u,\sigma^\text{MB}_i)\Theta(v)\,,
    \label{prob_v_prior}
\end{equation}
where $\bm{\theta}=\{T, w, c_0, c_1,\cdots,c_S\}$ are the unknown parameters to be inferred and $c_i=n_i/n$.

We model the collisions between individual gas particles and the nanoparticle as fully inelastic.
For a gas particle of mass $m_i$ moving at speed $v$, the recoil imparted on the nanoparticle is $p_\text{I}=\mu_iv$, for reduced masses $\mu_i=m_i M/(m_i+M)$, with probability density\footnote{We have chosen $p_\text{I}$ to be momentum of the nanoparticle only, excluding the adsorbed gas atom, and we note that since $m_i\ll M$, then $\mu \approx m_i$. This facilitates generalization to partially inelastic scattering, which we treat via a simple model in Appendix~\ref{sec:elastic_model}. There we find that partial inelasticity leads to a broadened recoil probability distribution.}
\begin{equation}
    \text{prob}(p_{\text I}|v,\bm{\theta}) = \sum_{i=0}^{S-1} c_i\delta(p_\text{I}-\mu_i v)\,.
    \label{prob_mom} 
\end{equation}
Marginalising Eq.~\eqref{prob_mom} over the speed $v$ with Eq.~\eqref{prob_v_prior}, we find the momentum probability distribution\footnote{We suppress the dependence of the probability distribution on the background information which includes the knowledge of the model and known parameters. We maintain this notation throughout this section.}
\begin{equation}
    \text{prob}(p_\text{I}|\bm{\theta}) = \sum_{i=0}^{S-1} c_i\frac{p_\text{I}}{\mathcal{P}_i}\mathcal{N}(p_\text{I}-\mathcal{P}_i,\sigma_i)\Theta\left(\frac{p_\text{I}}{\mu_i}\right)\,,
    \label{eq:recoil_distribution}
\end{equation}
where $\mathcal{P}_i=\mu_i u$, $\sigma_i=\mu_i\sigma^\text{MB}_i$.

We model the process of estimating $p_\text{I}'$ as Gaussian broadening over the true recoil momentum $p_\text{I},$ that is, $\text{prob}(p_\text{I}'|p_\text{I})=\mathcal{N}(p_\text{I}'-p_\text{I},\sigma_\text{det}),$ with $\sigma_\text{det}=3.15$ u km/s as identified in Sec.~\ref{sec:simulation}.
Then the probability distribution of the estimate $p_\text{I}'$ of $p_\text{I}$ is
\begin{align}
    &\text{prob}(p_\text{I}'|\bm{\theta}) = \int_{-\infty}^{\infty}\text{prob}(p_\text{I}'|p_\text{I})\text{prob}(p_\text{I}|\bm{\theta})\text{d}p_\text{I} \nonumber \\
    &=\sum^{S-1}_{i=0} c_i\frac{\alpha_i}{\mathcal{P}_i}\mathcal{N}\left(p_\text{I}' - \mathcal{P}_i,\sigma^\text{tot}_i\right)\left(2-\overline{\gamma}\left[-\frac{1}{2},\frac{\alpha_i^2}{2\beta_i^2}\right]\right)\,,
    \label{eq:measurement-process}
\end{align}
where
\begin{align}
    \sigma^\text{tot}_i &=\sqrt{\sigma_\text{det}^2+\sigma_i},
    ~~~
    \alpha_i=\frac{\sigma_i^2 p_\text{I}' +\sigma_\text{det}^2\mathcal{P}_i}{(\sigma^\text{tot}_i)^2},
    ~~~
    \beta_i=\frac{\sigma_\text{det}\sigma_i}{\sigma^\text{tot}_i}\,, \nonumber
\end{align}
and $\overline{\gamma}$ is the regularised incomplete gamma function.

We only record the recoil events with momentum above the threshold $p_\text{th}.$ Thus, for each detected event, we have the likelihood function\footnote{In principle, there should also be a higher momentum cut-off. There is a hard cut-off at the momentum which provides enough energy for the nanoparticle to escape the trapping potential. About $\sim 10^5$ u km/s would be required to allow a nanoparticle with $R=50$ nm to escape a $10^4$ K = 0.86 eV trap. The magnitude of the trap depth is proportional to the trap's local laser intensity \citep{dissertation-2117-95281}, and can be adjusted experimentally.
However, the relevant momenta in this work are not expected to approach this limit for typical experimental parameters used to trap nanoparticles, but further applications of LEVITAS to detect dust particles might need to consider and optimise this.}
\begin{equation}
\ell(p_\text{I}'|\bm{\theta}) \propto \text{prob}(p_\text{I}'|\bm{\theta})\Theta(p_\text{I}'-p_\text{th})\,.
\label{eq:likelihood_1}
\end{equation}
Assuming a collection of $N$ independent observations $\{p_\text{I}'\}_N=\{(p_\text{I}')_0,\cdots,(p_\text{I}')_N\}$ made in a duration $\tau$ we have
\begin{equation}
\ell(\{p_\text{I}'\}_N|\bm{\theta}) = \prod_j^N\ell([p_\text{I}']_j|\bm{\theta})\,.
\label{eq:likelihood_N}
\end{equation}
We use Eq.~\eqref{eq:likelihood_N} to perform maximum-likelihood estimation; to find maximum-likelihood estimates $\bm{\theta}_\text{MLE}$ for a given data-set $\{p_\text{I}'\}_N$, we maximise the logarithm of Eq.~\eqref{eq:likelihood_N} under the constraint $\sum_i c_i=1$, normalising Eq. \eqref{eq:likelihood_1} in $p_\text{I}'$ numerically at each evaluated $\bm{\theta}$. The confidence intervals corresponding to the best estimates can be constructed using the likelihood-ratio test \citep{Wilks_1938} by adjusting the parameter of choice in $\bm{\theta}$ while maximising Eq. \eqref{eq:likelihood_N} for all the others until the desired level of confidence is reached.

The best estimates of the composition weights $c_i$ can be used to estimate the number density corresponding to each species via $n_i=n' c_i$, where $n'=(N/\tau)(u \pi R^2)^{-1}$ is the estimate of the total number density $n$\footnote{Based on uncertainty propagation in the estimate $n'$, the counting statistics is expected to be the dominant source of error in the total number density over the calibration error due to the uncertainty in particle size when $N(\Delta R)^2/R^2<1$. Assuming a batch of nanoparticles with $ R=50\pm4$ nm, the uncertainty in the radius gives a contribution $N(\Delta R)^2/R^2 = N/625$. This calibration error could be lowered further by performing \textit{in-situ} measurements of the nanoparticle size using the measured nanoparticle position signal \citep{quidant_2019,Yu_2020}.}.
If the speed distribution is significantly truncated by the threshold $p_\text{th}$ (that is the number of missed events is larger than the error due to counting statistics $\sim \sqrt{N}$ for $N$ events measured in time $\tau$), which can be the case when very low momenta species (such as hydrogen) makes up a large fraction of the composition, it may be necessary to first estimate the missing fraction using the best estimates and Eq.~\eqref{eq:measurement-process} to accurately estimate the total number density. We estimate the missing fraction by computing $f_\text{missing}=\int_0^{p_\text{th}}\text{prob}(p_\text{I}'|\bm{\theta}_\text{MLE})\text{d}p_I'$ (See Eq. \eqref{eq:measurement-process}), which then gives us the collision rate $r'=(1-f_\text{missing})^{-1}(N/\tau)$ for $N$ observed impacts during time $\tau$, which we then use to estimate the total number density.
\section{Results} 
\label{sec:results}

\begin{figure*}
    \centering
\includegraphics[width=\linewidth]{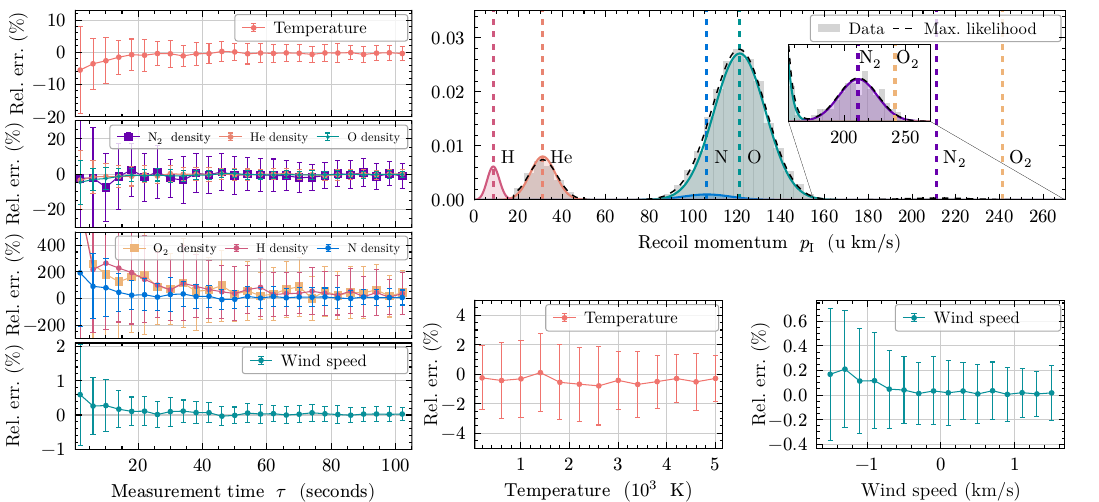}
\caption{Inference of parameters $\bm{\theta}=\{T,w,c_0,\cdots,c_5\}$ in LEO at 600 km altitude with $n=2.71\times 10^6/\text{cm}^3$ with 2.01\% H, 11.5\% He, 1.97\% N, 83.2\% O, 1.33\% $\text{N}_2$, and 0.13\% $\text{O}_2$.
    The error-bars shown represent sample standard deviation for 100 repeated runs of the simulation. In the case of wind speed, the relative error is taken w.r.t. the spacecraft speed $|v_s|=7.5$ km/s. \textbf{Left}: Variation of the relative error in the best estimates $\bm{\theta}_\text{MLE}$ with measurement time $\tau$ at $T=1045$ K and $w=0$ m/s. \textbf{Top right}: Example data histogram at $\tau=60$ s, $T=1045$ K, and $w=0$ m/s. The black dashed curve shows the Eq.~\eqref{eq:measurement-process} evaluated at $\bm{\theta}_\text{MLE}$, while the coloured curves with shaded areas correspond to the different terms contributing to the sum in Eq.~\eqref{eq:recoil_distribution} evaluated at $\bm{\theta}_\text{MLE}$. \textbf{Bottom right}: Relative error for different values of $T$ (middle) and $w$ (right), at $\tau=$ 60 s.}
    \label{fig:leo600}
\end{figure*}

We now illustrate the performance of LEVITAS in estimating the temperature, wind speed, and composition by simulating its operation for realistic sets of parameters corresponding to three low density scenarios. In Section \ref{sec:600}, we analyse a LEO regime at 600 km altitude to represent the uppermost altitudes at which conventional mass spectrometers might operate. In Section \ref{sec:1000}, we choose a LEO altitude of 1000 km which is instead targeted by remote observations of Lyman-alpha emission. Finally, in Sec. \ref{sec:interstellar}, we illustrate how neutrals from the interstellar medium observed can be detected by a spacecraft located beyond the Earth's magnetosphere at a Lagrange point.  
In order to derive moments of the underlying distribution functions, we assume that the species in these scenarios follow the Maxwell-Boltzmann distribution in one dimension over the measurement time $\tau$. 
These assumptions are made here for illustrative 
purposes. See Sec.~\ref{sec:disc} on how LEVITAS can provide a histogram of the raw particle impacts to resolve non-standard or multiple overlapping distribution functions.

For the highest number density considered in this section ($2.71\times 10^{6}/\text{cm}^3$ in LEO at 600 km altitude), the impact rate $n\pi R^2u$ for the parameters in Table~\ref{table:sim_params} is about $\lambda_0=$ 160/s. Assuming the impacts to be Poisson distributed with an average rate of $\lambda_0$, the probability of two or more impacts occurring in a time interval of the trap cycle
$P(k\geq2)\ll1\%$. Each impact can thus be unambiguously resolved in time as in Sec.~\ref{sec:tracking}.

\begin{figure*}
    \centering
\includegraphics[width=\linewidth]{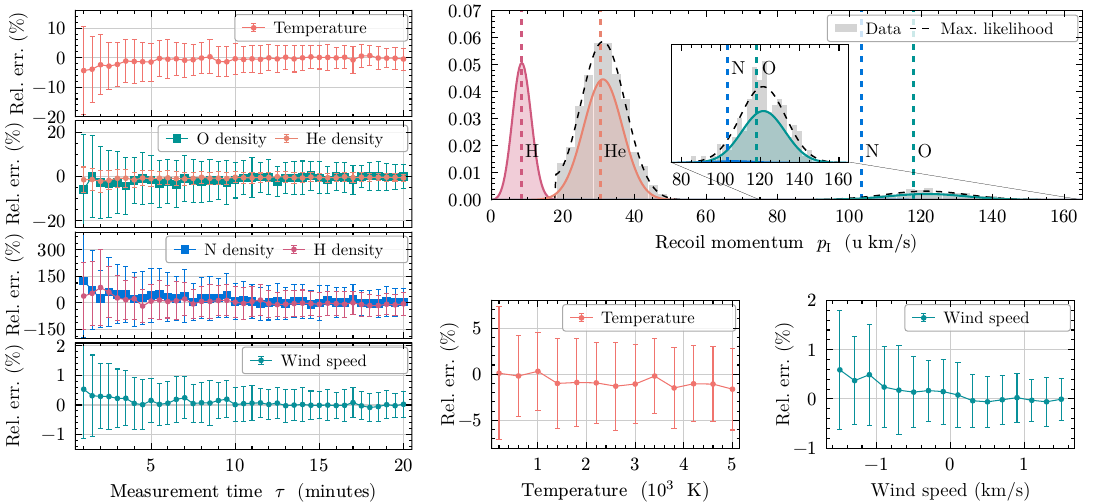}
\caption{Inference of parameters $\bm{\theta}=\{T,w,c_0,\cdots,c_3\}$ in LEO at 1000 km altitude with $n=1.21\times 10^5/\text{cm}^3$ with 31.6\% H, 61.9\% He, 0.3\% N, and 6.17\% O. The error-bars shown represent the sample standard deviation for 100 repeated runs of the simulation. In the case of wind speed, the relative error is taken w.r.t. the spacecraft speed $|v_s|=7.3$ km/s. \textbf{Left}: Variation of the relative error in the best estimates $\bm{\theta}_\text{MLE}$ with measurement time $\tau$ at $T=1045$ K and $w=0$ m/s. \textbf{Top right}: example data histogram at $\tau=300$ s, $T=1045$ K, and $w=0$ m/s. The black dashed curve shows the Eq. \eqref{eq:measurement-process} evaluated at $\bm{\theta}_\text{MLE}$, while the coloured curves with shaded areas correspond to the different terms contributing to the sum in Eq. \eqref{eq:recoil_distribution} evaluated at $\bm{\theta}_\text{MLE}$. \textbf{Bottom right}: Relative error for different values of $T$ (middle) and $w$ (right), at $\tau=$ 600 s.}
    \label{fig:leo1000}
\end{figure*}

To simulate the measurement process with LEVITAS, we sample the true recoil distribution in Eq.~\eqref{eq:recoil_distribution}, with additive Gaussian noise with variance $\sigma_\text{det}^2$ and truncated below the measurement threshold $p_\text{th}.$
We then use the procedure described at the end of Sec.~\ref{sec:gas_dynamics} to estimate the thermodynamic quantities.
The distribution parameters at 600 km and 1000 km (for Figs.~\ref{fig:leo600} and   \ref{fig:leo1000} respectively) are illustrative and obtained using NRLMSIS-2.1~\cite{emmert2022nrlmsis}  via the NASA Community Coordinated Modeling Center available at \url{https://kauai.ccmc.gsfc.nasa.gov/instantrun/nrlmsis/} for 15/07/2025, 00:00 UTC, 55\textdegree N, 45\textdegree E.

\subsection{LEO at 600 km} \label{sec:600}

At 600 km, for the parameters in Table~\ref{table:sim_params}, impacts occur at a rate about $\lambda_0=$ 160/s. This means that the medium is rare enough to resolve individual impacts, but dense enough for the quantities of interest to be estimated with good precision using a single nanoparticle probe within a few seconds of observation time. 

Our simulation results are shown in Fig.~\ref{fig:leo600}. The example histogram in top-right illustrates the estimated composition based on the maximum likelihood estimates, this is dominated by atomic oxygen at about 80\% of the total number density for this parameter set. For a given $\tau$, the achievable precision is higher for more abundant species, with only about 5 seconds needed to estimate the atomic oxygen number density with under $\pm$5\% precision.

Figure~\ref{fig:leo600} also shows the performance after $\tau = 60$ s for temperatures and wind speeds in the range $200\text{ K} \leq T \leq5000\text{ K}$ and $-1.5\text{ km/s} \leq w\leq1.5\text{ km/s}$, respectively. The wind speeds cause collective shifts of the peaks in $p_\text{I}$. The minimum measurable shift due to wind is determined by the width of the peaks. Thus, the temperature ultimately limits the precision of estimating the wind speed. In Appendix~\ref{sec:estimators}, we provide simple maximum likelihood estimators that capture the essence of this numerical procedure in a simplified scenario.

\subsection{LEO at 1000 km} \label{sec:1000}

For our chosen parameter set at 1000 km, the number density is about an order magnitude lower than at 600 km. The corresponding  impact rate is about $\sim 7$/s. 
This leads to measurements over longer time durations, as is evident in Fig.~\ref{fig:leo1000}.
$\tau \sim$~1~minute is required to constrain the full distribution of helium (which is the most abundant for this particular parameter set) with about $\pm$5\% precision. The measurement duration may be reduced by combining data from multiple sensor heads recording simultaneously; see Sec.~\ref{subsec:mux}. 

This scenario also highlights an aspect of our estimation strategy: the choice of threshold $p_\text{th}$.
With $p_\text{th}=18$ u km/s, a large fraction of the collisions corresponding to hydrogen are not observed, as shown by the hydrogen peak corresponding to the true distribution in the histogram in the top right of Fig.~\ref{fig:leo1000}. 
However, we find that even with significant truncation of the distribution, densities corresponding to abundant species can be accurately estimated as long as we account for the missing fraction of the data in the estimate of the collision rate $r'$ before estimating the total number density, as described at the end of 
Sec.~\ref{sec:gas_dynamics}. We discuss the threshold more generally and strategies to mitigate its limitations in Sec.~\ref{sec:disc}.

\begin{figure*}
    \centering
\includegraphics[width=\linewidth]{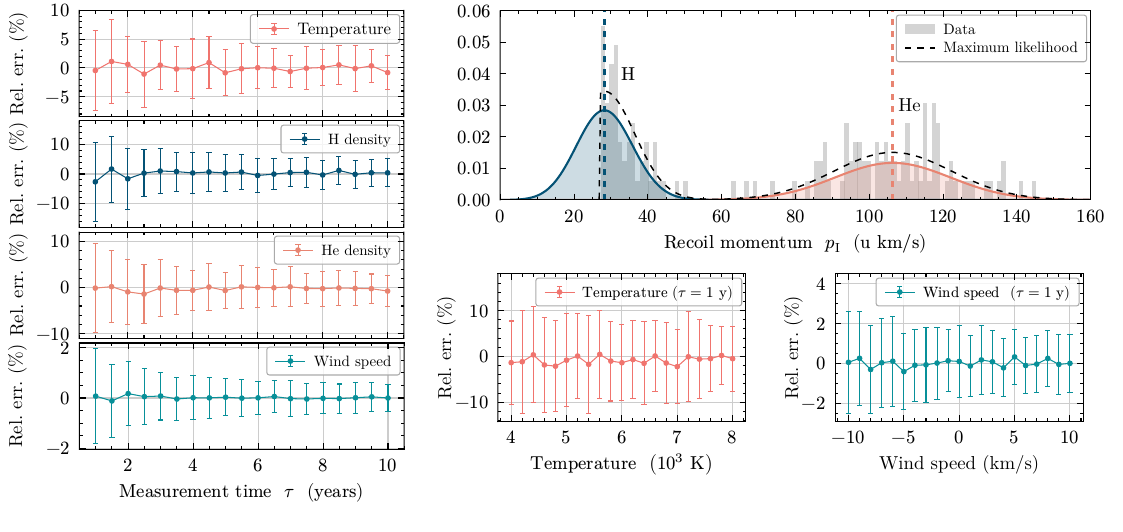}
    \caption{Inference of parameters $\bm{\theta}=\{T,w,c_0,c_1\}$ of the interstellar medium observed at L$_2$: $n=0.03/\text{cm}^3$ with 50\% $\text{H}$, and 50\% He \citep{bzowski2008,quemerais2009,Saul2012,witte2004}. The error-bars shown represent the sample standard deviation for 100 repeated runs of the simulation. In the case of wind speed, the relative error is taken w.r.t. the spacecraft speed $|v_s|=26$ km/s. \textbf{Left}: Variation of the relative error in the best estimates $\bm{\theta}_\text{MLE}$ with measurement time $\tau$ at $T=7500$ K and $w=0$ m/s. \textbf{Top right}: example data histogram at $\tau=1$ y, $T=7500$ K, and $w=0$ m/s. The black dashed curve shows the Eq. \eqref{eq:measurement-process} evaluated at $\bm{\theta}_\text{MLE}$, while the coloured curves with shaded areas correspond to the different terms contributing to the sum in Eq. \eqref{eq:recoil_distribution} evaluated at $\bm{\theta}_\text{MLE}$. \textbf{Bottom right}: Relative error for different values of $T$ (middle) and $w$ (right), at $\tau=$ 1 y. }
    \label{fig:L2Interstellar}
\end{figure*}

\subsection{Interstellar medium}
 \label{sec:interstellar}

For the last case study, we consider the possibility of detecting constituents of the very local interstellar medium \citep{Swaczyna_2022} {that have penetrated the heliosphere}. For this, we consider a probe located at the Earth--Sun Lagrange point 2 (L$_2$), where LEVITAS can benefit from a large orbital velocity for detection at an increased collision rate. Specifically, the data for this study was generated with spacecraft speed $|v_s|=26$ km/s. 

We perform our simulations with $n=0.03/\text{cm}^3,$ and assume an equal mix of hydrogen and helium to compare detecting the two species. This gives an average collision rate $\sim 0.5$/day. We set the threshold $p_\text{th} = 27 $ u km/s for which, assuming Gaussian white noise,\footnote{For Gaussian white noise with variance $\sigma^2$, $\text{prob}(p>p_\text{th})=\int_{-\infty}^{p_\text{th}}G(p,\sigma)\text{d}p=0.5\text{ erfc}(p_\text{th}/(\sqrt{2}\sigma))$ for each sample. For 1 MS/s and the noise floor shown in Fig.~\ref{fig:kick-demo}, $p_\text{th}=27\text{ u km/s}$ gives an expected rate of false detections of 0.05/day.} the expected number of false detections makes up $<10\%$ of all detected events. About 20\% of the recoil distribution corresponding to hydrogen is truncated below this threshold level.

Our simulation results are shown in Fig.~\ref{fig:L2Interstellar}. The number densities of both species can be estimated with less than $\pm 10$\% precision within about 1 year of total observation time with a single nanoparticle. This is an extreme case that requires long measurement times. However, we again note that with multiplexed detection across many nanoparticles, the total measurement time required can be reduced significantly. Multiplexing with, for instance, 100 nanoparticles reduces this to less than 4 days of measurement, as discussed in Sec.~\ref{subsec:mux}.

\section{Architecture of LEVITAS }
\label{sec:design}

\begin{figure}
    \includegraphics[width=0.5\textwidth]{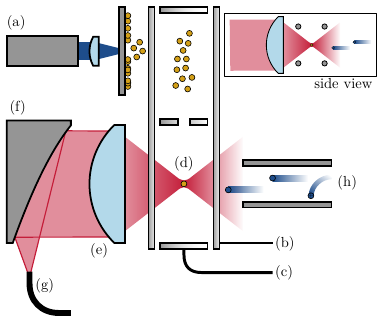}
    \caption{LEVITAS sensor head: Nanoparticles are ejected from a metallic substrate using a pulsed laser via laser-induced acoustic desorption (a), whereupon they are captured into a linear Paul trap comprising oscillating electrodes (b) and low-voltage endcaps (c).  The inset side view shows the circular cross-section of the four rod electrodes.  Several endcaps along the length of the linear Paul trap allow the cloud to be localized within and shuttled between segments, and the architecture includes upper ``capture'' and lower ``sensing'' regions. The captured nanoparticle cloud is moved to the sensing region where a single trapped nanoparticle (d) is captured into the optical trap, and the remainder returned to the capture region for later use.  The optical trap is formed by a high-aperture focusing lens (e) using light collimated by an off-axis parabolic mirror (f) provided by a single-mode optical fibre (g), connected via an optical circulator to both trapping laser source and homodyne detection. Fast moving gas atoms enter (h) and charged particles are removed from the stream using  electrostatic deflector plates.  Top-right: The side view is also shown in Fig.~\ref{fig:levitas_demo}, sans the electrodes.} 
    \label{fig:design}
\end{figure}

Having illustrated the potential of LEVITAS in 
measuring impacts of individual particles in 
space, we now outline its architecture. The LEVITAS sensor head comprises three subsystems: the nanoparticle delivery mechanism, the optical trapping and detection system, and control electronics, as depicted in Fig.~\ref{fig:design}. These can be integrated  within a compact, space-qualified package.

\subsection{Nanoparticle delivery system}

The nanoparticles are delivered using laser-induced acoustic desorption (LIAD), which ejects these particles from a solid substrate into the trapping region~\citep{nikkhou2021direct,bykov2019direct}.
Other approaches, such as those using MEMS devices have also been developed~\citep{khorshad2025invacuum}. Thermal absorption of a pulsed laser creates localized acoustic waves that propagate through the substrate and overcome surface adhesion forces. A single substrate can provide sufficient numbers of nanoparticles for LEVITAS to be deployed on extended missions.

The nanoparticles are initially captured into a radio-frequency (RF) Paul trap that offers a large capture volume, before being transferred to the optical trap for active detection. The Paul trap design follows established approaches from the trapped-ion community~\citep{schulz2006optimization}, enabling controlled particle transport and reliable loading into the optical detection region.

\subsection{Optical system implementation}
\label{subsec:optsys}

The optical trapping system uses high numerical aperture-focusing to achieve the tight confinement required for quantum-limited position detection. The trap geometry accommodates forward-facing operation with the trapping region exposed to the incoming neutral particle flux. A compact optical design maintains the nanoparticle at the focus while collecting backscattered light for interferometric position readout.

The back-scattered light enables phase-stable homodyne detection for position readout. Three-dimensional tracking can also be achieved through various approaches, including local multi-element detector configurations or fiber-coupled schemes using photonic lanterns~\citep{leonsaval2013photonic} or multi-core fibers that separate the collected light into multiple channels providing position information along different axes.

\subsection{Electronics and control architecture}

Our architecture separates the compact sensor head from the main electronics package. The sensor head, which contains only the optical assembly, nanoparticle delivery system, and Paul trap electrodes, connects via cables to the primary electronics housing the lasers, photodetectors, control electronics, and RF drive electronics. The control system implements real-time feedback using field-programmable gate array (FPGA) processing~\citep{setter2018realtime} for impact detection and autonomous operation with onboard data processing.

Power consumption is minimized through duty-cycle management and efficient laser operation. The electronics package includes radiation-hardened components, with the complete instrument designed for integration on small satellite platforms.

\section{Extensions to LEVITAS}
\label{sec:extensions}

The LEVITAS sensor head described above is a minimal design that can be modified and augmented in multiple ways to extend its performance. These include the ability to function at higher densities, increasing the event rate and improving sensitivity to lower momentum impacts. Ultimately, we suggest the implementation of quantum sensing techniques exploiting non-classical states of the nanoparticle motion and probe light to go below the standard quantum limit.

\subsection{Pressure Mode}

At high gas densities, the collision rate may exceed the temporal resolution for detecting individual events. In this regime, LEVITAS operates in a complementary ``pressure mode'' where the sensor transitions from the microcanonical detector of individual impacts to ensemble measurement.

In the pressure mode, the continuous bombardment of gas particles produces two measurable effects. First, the time-averaged momentum flux creates a static displacement of the nanoparticle from its equilibrium position, proportional to the ram pressure. Second, statistical fluctuations in the collision rate manifest as enhanced Brownian motion, effectively increasing the thermal noise spectrum of the oscillator, which could be analyzed to extract collision statistics.

The transition to the pressure mode can be controlled dynamically by adjusting the detection threshold $p_\text{th}$ and averaging window $\Delta$, allowing adaptive operation in media across a wide range of densities. The density at which the transition is performed can be set to allow some margin before the impact rate exceeds the trap frequency $\Omega$. However, LEVITAS offers multiple trade-offs as to when this actually happens. These include
the nanoparticle radius $R$, on which the impact rate depends as $\propto R^2$, the optical trap depth depends as $\propto  R^3$, and minimum momentum for individual events that can be detected, which depends as $\propto R^{3/2}$. The last is the standard quantum limit (SQL) addressed in Sec.~\ref{sec:qenhance}. 
Laser power $P$ affects trap frequency $\Omega\propto P^{1/2}$ and thereby also SQL momentum resolution $\propto P^{1/4}$.
Nanoparticle size must remain sub-wavelength $R\ll \lambda$, and the fast reduction in both optical trap depth and scattered field amplitude for smaller particles (both $\propto R^3$) suggests increasing laser power to compensate, with this increase limited by nanoparticle material properties and the satellite power budget.

\subsection{Multiplexing}
\label{subsec:mux}

Our inference strategy in Sec.~\ref{sec:gas_dynamics} is such that the statistical precision of the estimates improves with observation time. It can lead to prohibitive measurement times in extremely rare media, such as in Fig.~\ref{fig:L2Interstellar}. This can be mitigated by monitoring multiple nanoparticles, thereby increasing the collection area. 
Several approaches exist for implementing this, with varying degrees of complexity~\citep{melo2024vacuum,siegel2025optical}.

The simplest architecture employs multiple sensor heads connected to a shared central control unit, with common laser sources, high-voltage supplies, and control electronics. 
More compact implementations can accommodate multiple nanoparticles within a single sensor head, separated by only millimetres.
A more sophisticated approach uses a two-dimensional optical lattice orthogonal to the ram direction~\citep{siegel2025optical}. Counter-propagating optical fields create a standing wave array with nanoparticles captured at antinodes. Separate readout beams allow for detection of momentum kicks along different directions.

A modest $10\times10$ array of $\sim 100$ nanoparticles increases the effective observation time by a factor of $\sim 100$. For the interstellar medium measurements discussed in Sec.~\ref{sec:interstellar}, this reduces the required measurement time from $\sim 1$ year to less than 4 days for the same statistical precision. The multiplexing factor can be adjusted adaptively: high-density environments benefit from continuous monitoring of fewer particles, while rarefied environments require many particles to achieve sufficient impact statistics.

\subsection{Nanoplates}

Thus far, we have considered spherical nanoparticles for which collisions occur randomly across the projected surface. As discussed in Appendix~\ref{sec:elastic_model}, this geometric randomness combined with elastic scattering produces a broadened recoil momentum distribution that limits the precision with which thermodynamic properties can be inferred.

An alternative is to use flat nanoplates oriented perpendicular to the ram direction~\citep{winstone2022optical} where the collisions have nearly collinear momentum transfer, eliminating the geometric broadening that affects spherical particles.

Orientation control is achieved through optical torque from the anisotropic polarisability of the flat geometry. When illuminated by a linearly polarized trapping beam, the plate experiences a restoring torque for angular deviations from alignment with the beam polarization. For a disc with a 1:10 aspect ratio and volume comparable to a 100 nm sphere, experimental work~\citep{winstone2022optical,gao2024feedback} and calculations based on ellipsoidal polarisability models~\citep{bohren1998absorption} suggest libration frequencies in the hundreds of kHz range, enabling strong orientational confinement.

\subsection{Three-dimensional momentum resolution}

Our work has focussed on one-dimensional tracking along the ram direction.
Three-dimensional tracking of recoil trajectories of the nanoparticle would enable significant extensions to the capabilities of LEVITAS. As noted in Sec.~\ref{subsec:optsys}, this can be achieved through multi-element detector configurations or fiber-coupled schemes that separate collected light into multiple channels.

The augmented capability is of resolving the wind \emph{vector} rather than just the ram-direction component. This provides full velocity field information for atmospheric dynamics studies or targetting energetic neutral atoms emission at the heliopause in planetary environments.
Three-dimensional tracking would also enable empirical determination of the coefficient of restitution $\alpha$. As $\alpha$ increases from zero, recoils expand from purely anti-ram to a cone with $\alpha$-dependent half-angle; see Appendix~\ref{sec:elastic_model}.

\subsection{Optical geometries and trapping potentials}

In this work, we have considered the canonical optomechanical setup of using a single-beam and high numerical aperture for trapping and light collection. However, there is considerable design freedom to modify the optical trapping potential of the nanoparticle to operate in different sensing regimes. Light collection can also be decoupled from the trapping beam, expanding the design freedom.

To improve sensitivity to low momentum transfer events, the trap ``tightness'' can be relaxed by engineering a low frequency $\Omega$ whilst still capturing a large fraction of the scattered light and maintaining a high intensity at the nanoparticle. This can be achieved through optical potential engineering~\citep{grimm2000optical} using structured light~\citep{forbes2021structured,yang2021optical}, alternative beam geometries~\citep{zhan2009cylindrical}, or multiple beam configurations. The reduced spring constant of the trapping potential allows for a greater excursion of the particle for a given momentum kick which results in a greater sensitivity to impacts. In the limit of low $\Omega$, this corresponds to the quasi-free particle regime; see Section \ref{sec:qenhance} for quantum non-demolition measurement of momentum in this regime.

The trapping potential does not have to be harmonic: it may be possible to exploit higher-order non-linear terms for different readout mechanisms, such as those based on side-band detection. This can be achieved with non-Gaussian laser beams, for example Bessel beams, box traps, or incoherent superpositions of Gaussian beams. Exploiting orbital angular momentum~\citep{simpson1997mechanical} or Poincar\'e beams~\citep{wang2012optical} is yet another possibility for the creation of different trapping potentials with attractive characteristics such as reduced heating rates~\citep{almeida2025levitated}.

\subsection{Multi-Channel Measurements}

Differently sized nanoparticles can be used in the same instrument across multiple sensor heads to provide complementary measurements. Smaller nanoparticles are more sensitive to momentum transfer events, but have a lower impact rate than larger particles. The size of the dispensed nanoparticle could be adapted to the measurement conditions (range of expected impact  momenta and density), or if multiple trapping regions are available, different nanoparticle sizes can be operated simultaneously and their data combined.

LEVITAS can also be flown together with other conventional accelerometers or mass spectrometers, or other quantum instruments such as cold atom interferometers for thermosphere drag measurements~\citep{Siemes2022} for cross-calibration.

\subsection{Quantum Enhancement}
\label{sec:qenhance}

Heisenberg's uncertainty principle limits the precision with which the momentum of a particle can be measured. This is given by the standard quantum limit (SQL) $p_{\text{SQL}}=\sqrt{\hbar M \Omega}$~\citep{Caves1980}.  It may thus be thought that this sets a lower limit to the momentum impacts that LEVITAS can detect. 
For the parameters in Table~\ref{table:sim_params}, $p_{\text{SQL}} \sim 1.88$ u km/s.

However, lower momentum impacts can be detected using techniques such as quantum mechanical squeezing and quantum non-demolition measurements. 
Squeezing can be applied to the quantum mechanical state of the motional degrees of freedom of the nanoparticle or the light probing the system, or both. The former have been produced experimentally ~\citep{Rashid2016,rossi2025quantum} and
suppress the uncertainty in position measurements. The nanoparticle in LEVITAS may be prepared in such a quantum state to measure the position more precisely, thus 
detecting smaller momentum impacts.
Squeezing the quantum state of the light is a standard technique used in gravitational wave detectors~\citep{Ganapathy2023}. Similar techniques are suggested to resolve lower momentum impacts in levitodyamic sensors~\citep{lee2025}.

The origin of the SQL lies in measuring the position which does not commute with the momentum at the level of quantum operators. It can thus be circumvented by measuring the momentum directly. One such possibility is to use a speedmeter~\citep{BRAGINSKY,Khalili1996}. It measures the difference of positions at two times directly by placing mirrors connected by an optical delay line (of time $t_d$) around the nanoparticle. The difference of its positions at two times separated by $t_d$ is read out interferometrically~\citep{Ghosh2020}, thus avoiding the SQL. 

Quantum enhancements may thus enable LEVITAS to detect impacts of very low magnitudes, albeit limited by optical losses and other imperfections. Nevertheless, it could be the only way forward at even higher altitudes than 1,000 km where the relative orbital velocity and low mass of hydrogen approaches the lower momentum threshold of LEVITAS. This might also be relevant when targeting neutral and dust populations in the environments of smaller solar system planets and bodies where orbital velocities are similarly reduced.

\section{Discussion}
\label{sec:disc}

We have presented the operating concept and architecture of LEVITAS, a novel levitodynamic sensor for accurate in-situ sensing of individual particles in space. We have illustrated its promise in three quite distinct scenarios.

Going forward, the challenge for LEVITAS lies in detecting smaller momenta impacts in rarer media. This leads to a trade-off in the choice of the momentum threshold $p_\text{th}$ and inferring false impacts. In denser media, a lower threshold level can be chosen as higher rates of false detections can be tolerated if there are more impacts overall. Future studies may enable detecting recoil events at a lower threshold level without a detrimental increase in the rate of false events, for instance, with the matched filter method \citep{Moore2024}, which also has a well developed framework for studying the rate of false detections \citep{Morras_2023,Vio_2017}. 

Detecting smaller momenta will require improvements, first classical and then quantum. 
However, there is an interplay between the lowest detectable momentum and the lowest density at which the sensor can operate, independent of the detection method used. This is because the mechanical noise driving the nanoparticle dynamics due to the scattered photons can cause sufficiently large recoil to mimic an expected impact from a neutral gas species. Integrating this noise over a time $t$ yields the rate of false detections $\text{erfc}(p_\text{th}/(\sqrt{2}\sigma_t))/2t$ where $\sigma_t=2\pi S_\text{FF}\sqrt{t}$ is the standard deviation of the Wiener noise process driving the dynamics. 
For the parameters in Table~\ref{table:sim_params}, setting $p_\text{th} = p_{\text{SQL}} \sim$ 1.88 u km/s and insisting on no more than 10\% false detection rate over an integration time of $t=1$ \textmu s leads to a minimum observable density $\sim 10^4/\text{cm}^{3}$ at typical LEO speeds. 
However, for the same parameters, setting $p_\text{th} = 2p_{\text{SQL}},$ the minimum observable density is practically nil. This is because $\sigma_t \sim $ 0.35 u km/s for the parameters in Table~\ref{table:sim_params} and for $t=1$ \textmu s, doubling the threshold is equivalent to displacing it by about six standard deviations.

Future studies could extend our inference methodology to a sensor operating in three spatial dimensions. This will enable the estimation of wind velocity vectors, leveraging the three-dimensional tracking of the position of the nanoparticle. It could also provide insights into the scattering mechanism of the species off the nanoparticle, spanning the extremes between perfectly elastic and fully inelastic collisions.

Our inference methodology could also be extended to scenarios where the neutral gas particles are not Maxwell-Boltzmann distributed. This is of significance in the study of oxygen populations in the upper atmosphere that are out of local thermal equilibrium \citep{Sharma94}, charged populations with high energy tails \citep{Pierrard2010}, or multiple unknown or overlapping distribution functions \citep{Rhea}. The histogram of the magnitude of the recorded impacts from LEVITAS along with their time stamps can provide the raw data, from which methods to calculate moments can be refined or customised.

\begin{acknowledgments}
This work has been supported by European Space Agency Contract No. 4000145824/24/NL/FFi, UK Space Agency grant ETP1-025 LOTIS, 
EPSRC International Network in Space Quantum Technologies INSQT (EP/W027011/1),
EPSRC Integrated Quantum Networks Hub (EP/Z533208/1) and EPSRC Quantum Technology Hub in Quantum Communications (EP/T001011/1).
RTD acknowledges a Science and Technology Facilities Council Ernest Rutherford Fellowship ST/W004801/1 and UKSA grant ST/Y005635/1.
AD acknowledges the UKRI “Quantum
Technologies for Fundamental Physics” programme
(Grant Numbers ST/T006404/1, ST/W006308/1 and
ST/Y004493/1). BE acknowledges the
EPSRC (Grant Numbers EP/R004773/1 and EP/M009386/1).
We thank Olivier Carraz for fruitful discussions. 
\end{acknowledgments}
  
\appendix

\section{Nanoparticle $T_\text{CM}$ requirement} \label{sec:Tcm_reqs}

The spring constant of a nanoparticle trapped in an optical trap is proportional to the focal laser intensity, making the position of the nanoparticle susceptible to fluctuations in the laser power~\citep{savard1997heating}. This is a parametric heating mechanism that adds energy to the oscillator with angular frequency $\Omega$ and energy $E$ at a rate
\begin{equation}
    \dot{E}=\frac{\pi}{2}\Omega^2S_\text{RIN}(2\Omega)E
\end{equation}
where $S_\text{RIN}$ is the spectral density of the laser relative intensity noise. In high vacuum, when gas damping is no longer the dominating damping mechanism, this laser intensity noise makes feedback cooling of the nanoparticle motion necessary for stable trapping \citep{Jain_2016}. The contribution of laser intensity noise to heating becomes negligible when the c.m. motion is cooled below a temperature $T_\text{CM}$ at which noise due to laser-induced backaction dominates, which adds energy to the oscillator at a rate $\dot{E}=\pi S_\text{FF}/M$, \citep{gajewski_2024} with $S_\text{FF}$ given by Eq.~\eqref{eq:force-spectral-density}. The two heating mechanisms contribute equally when
\begin{equation}
    T_\text{CM} = \frac{2S_\text{FF}}{S_\text{RIN}(2\Omega)}\frac{1}{M\Omega^2k_B}
\end{equation}
where $k_B$ is the Boltzmann constant. For the parameters listed in Table~\ref{table:sim_params} and a laser with typical $S_\text{RIN}(2\Omega)=-130\text{ dB/Hz}$~\citep{fu2017review}, gives $T_\text{CM}=24$ K. Therefore, demanding $T_\text{CM}\sim1$ K renders the laser intensity noise negligible; this is easily achieved~\citep{gieseler2012subkelvin}.

\section{Kalman filter details} \label{sec:kalman_filter}
To track the motion of the nanoparticle and make estimates of the collision impulses, we use a discrete Kalman filter \citep{Lessard_2021_11}. We track the nanoparticle state in one dimension and the collision force, based on measurements of the position. Therefore, the state vector used in the Kalman filter is given by $\mathbf{s}_n=(z_n,(v_z)_n,(F_\text{I})_n)$, where $n$ denotes the time-step, and $\mathbf{H}=(1,0,0)^\top$ the observation vector.  The dynamics model for the nanoparticle state is that of a damped harmonic oscillator, extended with a constant acceleration model for the dynamics of the impulse force $F_\text{I}$; the transfer matrix $\mathbf{A}$ used to predict the state vector for the time step $t_{n+1}$ via\footnote{The hat here denotes the posterior state estimate, obtained with the Kalman update of the predicted prior state estimate.} $\mathbf{s}_{n+1}=\mathbf{A}\mathbf{\hat{s}}_n$  is given by,

\begin{equation}
    \mathbf{A}=
    \begin{pmatrix}
\cos(\Omega\Delta t) & \frac{1}{\Omega}\sin(\Omega\Delta t) & \frac{1}{2m}\Delta t^2\\
-\Omega\sin(\Omega\Delta t) & \cos(\Omega\Delta t) & \frac{1}{m}\Delta t\\
0 & 0 & 1
\end{pmatrix}
\end{equation}
where $\Delta t = t_{n+1}-t_n$ and we assumed that $\gamma\ll\Omega$ and $\gamma\Delta t\ll1$. The covariance matrix of the process noise is set to,

\begin{equation}
    \mathbf{Q}= \sigma^2_Q
    \begin{pmatrix}
\Delta t^3/3 & \Delta t^2/2 & 0\\
\Delta t^2/2 & \Delta t & 0\\
0 & 0 & \Delta t^{-1}
\end{pmatrix}
\label{process_noise}
\end{equation}
where $\sigma_Q^2$ is the spectral density of the force fluctuations. The top left $2\times2$ sub-matrix of \eqref{process_noise} corresponds to discrete process noise due to white noise \citep{Setter_2018}. The process variance for $F_\text{I}$ is set to $\sigma_Q^2/\Delta t$ to avoid tracking the white noise force causing the Brownian motion of the oscillator, and it is set uncorrelated with the other components. The measurement noise variance is set by imprecision due to shot-noise in a back-scatter detection system \citep{Tebbenjohanns2019}.

\section{Bayesian inference details} \label{sec:bayesian}
\begin{figure}
    \centering
    \includegraphics[width=\linewidth]{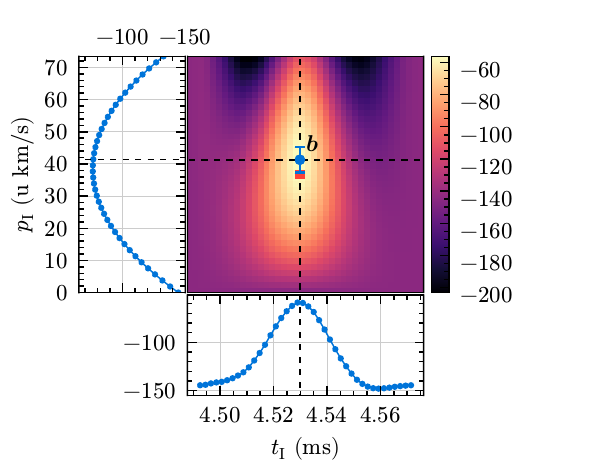}
    \caption{Example of the marginal log-posterior profile $\mathcal{L}(t_\text{I},p_\text{I})=\log(\text{prob}(t_\text{I},p_\text{I}|\{Z_i\},I))$ from Eq. \eqref{marginal-general} (not normalised), centered on a true collision event marked by the red square at $t_\text{I}$=4.53 ms and $p_\text{I}=$ 36.73 u km/s. The blue point $\mathbf{b}$ marks the maximum of $\mathcal{L}$ and the location of the best estimate in the $(t_\text{I},p_\text{I})$ domain. The bottom and left plots show the horizontal and vertical slices of $\mathcal{L}$ respectively, about $\mathbf{b}$.}
    \label{fig:marginal_profile}
\end{figure}
The posterior probability distribution is derived based on a dynamical model of harmonic motion, impulse response due to a single impact at time $t_\text{I}$ and Gaussian-distributed white noise $w(t)$ with variance $\sigma^2$,
\begin{equation}
   z(t,\bm{\theta}) = z_0\sin(\Omega t +\phi) + \frac{p_\text{I}}{M\Omega}\sin(\Omega[t-t_\text{I}])\Theta(t-t_\text{I}) + w(t)
   \label{dynamics_model_bayesian}
\end{equation}
where $\bm{\theta}=\{z_0,\phi,t_\text{I},p_\text{I}\}$ in which $z_0$ and $\phi$ are nuisance parameters, $p_\text{I}$ and $t_\text{I}$ are the impact time and the recoil momentum amplitude, respectively, and $\Theta$ is the Heaviside step function. The second term in Eq.~\eqref{dynamics_model_bayesian} represents the impulse response function corresponding to Eq.~\eqref{eom} with $\gamma\ll\Omega$ and evolution over a time $T\ll\gamma^{-1}$. The variance of the process driving the random-walk dynamics in the trajectory grows as $\sigma_Q^2t^3$ \citep{setter_2020}, where $\sigma_Q^2=2\pi S_\text{FF}/M^2$ is the variance of acceleration fluctuations. Therefore, the size of the data window used for analysis is chosen $\lesssim (\sigma/\sigma_Q)^{2/3}$, for which we expect the noise in the signal to be dominated by the measurement noise $w(t)$. 

We assume that our measurement gives us a direct readout of the nanoparticle displacement $Z_i$ at time $t_i$. For a collection of $N$ independent measurements $\{Z_i\}$, we obtain the likelihood function \citep{Sivia_2006},

\begin{equation}
\text{prob}(\{Z_i\}|\bm{\theta},I)=\left(\frac{1}{2\pi\sigma^2}\right)^{N/2}\prod_i^N\exp\left(-\frac{1}{2\sigma^2}[Z_i-z(t_i,\vec{\theta})]^2\right) 
\end{equation}
where $I$ represents the background information, which includes the dynamics model, the normal distribution of the noise, and the values of the known parameters $M,\Omega$. We use Bayes' theorem to write down the marginal posterior distribution, \citep{Sivia_2006}

\begin{equation}
    \text{prob}(t_\text{I},p_\text{I}|\{Z_i\},I) \propto \iint \text{prob}(\{Z_i\}|\bm{\theta},I)\pi(\bm{\theta},I)\text{d}\phi\text{d}z_0
    \label{marginal-general}
\end{equation}
with a uniform prior $\pi(\bm{\theta},I)$ for all parameters in $\bm{\theta}$. For the nuisance parameters $z_0$ and $\phi$, the prior distribution provides integral limits in the ranges $-Z\leq z_0<Z$ and $0\leq \phi < \pi$, respectively, for some positive $Z$. We approximate the marginal integral for $z_0$ with the analytical integral with limits extended to $\pm\infty$, and obtain the marginal log posterior as
\begin{equation}
\log(\text{prob}(t_\text{I},p_\text{I},\phi|\{Z_i\},I))=c+\frac{1}{2}\log\left(\frac{\pi}{A}\right) + \frac{B^2}{4A}-\frac{1}{2\sigma^2}\sum_{i=1}^N a_i^2
\label{marginal-phi}
\end{equation}
for a normalisation constant $c$, and
\begin{align*}
   a_i&=Z_i-\frac{p_\text{I}}{M\Omega}\Theta(t_i-t_\text{I})\sin(\Omega[t_i-t_\text{I}]) \\
   A&=\frac{1}{2\sigma^2}\sum_{i=1}^N\sin^2(\Omega t_i+\phi)\\
   B&=\frac{1}{2\sigma^2}\sum_{i=1}^Na_i\sin(\Omega t_i+\phi)\\
\end{align*}
To obtain the log of Eq.~\eqref{marginal-general} $\mathcal{L}(t_\text{I},p_\text{I})=\log(\text{p}(t_\text{I},p_\text{I}|\{Z_i\},I))$, we marginalise Eq.~\eqref{marginal-phi} numerically over $\phi$. To avoid numerical over-/under-flows, we employ the \texttt{lintegrate} library \citep{Pitkin_2022} to compute the integral in log-space. An example profile of the log marginal posterior for one flagged event is shown in Fig.~\ref{fig:marginal_profile}. The plot shows the profile in the neighbourhood of the initial guess obtained from the Kalman filter, and the best estimate $\mathbf{b}$ obtained by maximising the output of $\mathcal{L}$. The errors in the estimates of $t_\text{I}$ and $p_\text{I}$ are obtained using the diagonal elements of the Hessian matrix about the best estimate, \citep{Sivia_2006}
\begin{equation}
   \Delta t_\text{I} = \Bigg(-\frac{\partial^2\mathcal{L}}{\partial t_\text{I}^2}\bigg\vert_\mathbf{b}\Bigg)^{-1/2} \qquad
      \Delta p_\text{I} = \Bigg(-\frac{\partial^2\mathcal{L}}{\partial p_\text{I}^2}\bigg\vert_\mathbf{b}\Bigg)^{-1/2} 
\end{equation}
and represent the standard deviation of a Gaussian distribution.

\section{Modelling inelastic collisions} \label{sec:elastic_model}

Atom-surface scattering at thermal energies ($\sim k_B T \approx 0.025$~eV at 300~K) is well-described by phenomenological models involving an accommodation coefficient that parametrizes the mixture of specular (elastic) and diffuse (inelastic) reflection~\citep{cavalleri2010gas,martinetz2018gasinduced}.
This framework has been experimentally validated for nanomechanical vacuum gauges~\citep{blakemore2020absolute}.
However, gas-surface collisions in low-Earth orbit occur at hyperthermal energies (several eV, corresponding to spacecraft velocities of 7--8 km/s), where scattering dynamics is not adequately captured by existing gas-surface interaction models~\citep{jorge2025modeling}.
The details of momentum transfer at hyperthermal energies depend in a complex manner on the atomic species, surface properties, collision energy, and scattering geometry, constituting an active area of research.

For the present work, we adopt a simplified collision model parametrized by a coefficient of restitution $0\leq\alpha<1$, which provides a tractable description of momentum transfer.
High energy accommodation coefficients ($\alpha_E \sim 0.5$--$0.8$) observed for hyperthermal oxygen atoms on satellite materials~\citep{jorge2025modeling} indicate substantial energy transfer to the surface, corresponding to low coefficients of restitution.
We therefore focus on the fully inelastic limit ($\alpha=0$) in the main text. We treat the general partially inelastic case here.

We derive the recoil momentum distribution by first considering a single species of mass $m_i$, and later generalising to multiple species. As before, we consider the incoming atoms to be moving towards the trapped nanoparticle along the direction of motion of the spacecraft, with momentum $m_iv$. We focus the derivation on nanoparticle recoil momentum component along the spacecraft direction, which we denote $p_\text{I}$.

The atoms collide with the nanoparticle at a polar angle $\theta$ in the spherical coordinate system centered on the nanoparticle. We assume that $\alpha$ is constant and independent of $\theta$. By considering the projected area element of a sphere $\text{d}A \cos\theta =R^2\sin\theta\cos\theta\text{d}\theta\text{d}\phi$, we find that the probability that the collision occurs at $\theta$ is $\text{prob}(\theta)=2\sin\theta\cos\theta$ with $0\leq\theta<\pi/2$ and no dependence on the azimuthal angle.

\begin{figure}
    \centering
    \includegraphics[width=\linewidth]{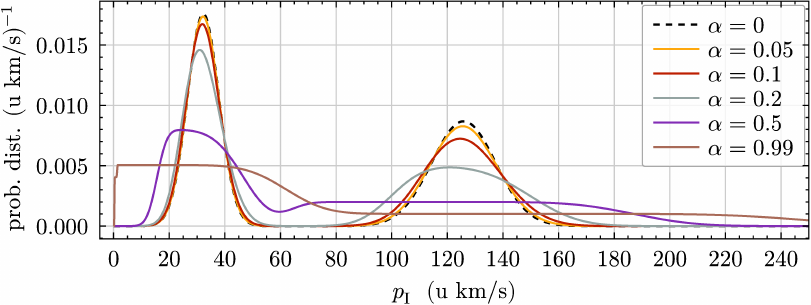}
    \caption{Example recoil momentum probability distribution in Eq.~\eqref{recoil_distribution_general} plotted for a range of $\alpha$ for equal proportion of helium and oxygen at $T=1000$ K, $w=0$ m/s, $M=$ 1 fg and $v_s=7.8$ km/s. The plot also shows the equivalent recoil distribution for the fully inelastic collisions (Eq.~\eqref{eq:recoil_distribution} from the main text), marked by the black dashed line.}
    \label{fig:prob_p_alpha_range}
\end{figure}

Fully inelastic ($\alpha=0$) collisions impart the same momentum recoil, independently of the location of impact on the nanoparticle. In Sec.~\ref{sec:gas_dynamics}, this is expressed as $\text{prob}(p_\text{I}|v,\bm{\theta})=\delta(p_\text{I}-\mu_iv)$. For partially inelastic collisions, the momentum recoil along the spacecraft direction is a function of the polar angle $p_\text{I}=(1+\alpha\cos2\theta)$, giving a possible range $m_i v(1-\alpha) \leq p_I < m_iv(1+\alpha)$. Changing variables \citep{Sivia_2006} to $p_\text{I}$ we find
\begin{align}
\label{eq:prob_elastic}
    \text{prob}(p_\text{I}|v,\alpha,\bm{\theta}) & =  \text{prob}(\theta|v,\alpha,\bm{\theta})\left|\frac{\text{d}\theta}{\text{d}p_\text{I}}\right| \\\nonumber
    &    = 
    \begin{cases}
    \dfrac{1}{2m_i\alpha v} & m_i v(1-\alpha) \leq p_I < m_iv(1+\alpha)\\
    0 & \text{otherwise}.  
    \end{cases} 
\end{align}
Eq.~\eqref{eq:prob_elastic} describes a uniform distribution over the full range of possible momentum recoil allowed by the geometry and the coefficient of restitution. Repeating the marginalisation procedure leading to Eq.~\eqref{eq:recoil_distribution}, we again use Eq.~\eqref{prob_v_prior} to find the recoil distribution
\begin{align}
    \text{prob}(p_I|\alpha,\bm{\theta}) &= \int_{-\infty}^\infty\text{prob}(p_I|v,\alpha,\bm{\theta})\text{prob}(v|\bm{\theta})\text{d}v\\
    &=\frac{1}{2\alpha m_i u}\int_{a_i(p_\text{I})}^{b_i(p_\text{I})}G(v-u,\sigma_i^\text{MB})\text{d}v\nonumber\\
    &=\frac{1}{4\alpha m_i u}\left(\text{erf}\left[\frac{b_i(p_\text{I})-u}{\sqrt{2}\sigma_i^\text{MB}}\right] - \text{erf}\left[\frac{a_i(p_\text{I})-u}{\sqrt{2}\sigma_i^\text{MB}}\right]\right)\nonumber
\end{align}
where
\begin{align}
   a_i(p_\text{I})=\frac{p_\text{I}}{m_i(1+\alpha)},
~~~   b_i(p_\text{I})=\frac{p_\text{I}}{m_i(1-\alpha)}.
\end{align}
For multiple species, this distribution generalises to
\begin{equation}
    \text{prob}(p_I|\alpha,\bm{\theta}) =\frac{1}{4\alpha u}\sum_i^S\frac{c_i}{m_i}\left(\text{erf}\left[\frac{b_i(p_\text{I})-u}{\sqrt{2}\sigma_i^\text{MB}}\right] - \text{erf}\left[\frac{a_i(p_\text{I})-u}{\sqrt{2}\sigma_i^\text{MB}}\right]\right).
    \label{recoil_distribution_general}
\end{equation}
The elasticity of the collision acts as an additional broadening mechanism for the recoil distribution as shown in Fig.~\ref{fig:prob_p_alpha_range}. The above distribution approaches Eq.~\eqref{eq:recoil_distribution} as $\alpha\rightarrow0$.

In practice, the value of $\alpha$ is not well known. For normal incidence in the hyperthermal limit, $\alpha \approx \sqrt{1-\alpha_E}$, suggesting $\alpha \sim 0.4$--$0.7$ for the observed $\alpha_E \sim 0.5$--$0.8$, though the value may vary depending on surface geometry, surface roughness, and the atomic species being scattered~\citep{jorge2025modeling}.
The problem can be reduced through marginalisation of Eq.~\eqref{recoil_distribution_general} in $\alpha$, with priors informed by the relationship to measured energy accommodation coefficients.
In principle, $\alpha$ can be determined empirically with three-dimensional momentum tracking. The angular distribution of recoil momenta depends on $\alpha$: if $\alpha=0$, all momentum transfers occur in the anti-ram direction; as $\alpha$ increases, the recoils spread over a cone with an $\alpha$-dependent half-angle. This characteristic angular signature, together with direct measurement of the kinetic energy transfer, enables direct measurement of the coefficient of restitution, particularly in controlled laboratory settings using monoenergetic atomic beams where the incident energy is well-defined.

\section{Fisher information and estimators}
\label{sec:estimators}
To assess the merit of the general approach to inferring gas properties from the relative momenta of gas particles sampled from the distribution and to benchmark the capabilities of the proposed LEVITAS instrument, we consider the best possible inferences that can be made \emph{in principle} from this information.
We do this by assuming the measurement noise for individual recoil momenta is negligible and by computing the total available Fisher information.  We further simplify by considering a single gas species of mass $m$ and temperature $T$.
For this simplified case, we define the inference parameters as $\bm{\theta}=\{T, u\}$, where $u = v_s + w$ is the relative speed between the spacecraft and the gas.

While the true velocity distribution of gas particles is Maxwellian $v \sim \mathcal{N}(u,\sigma^\text{MB})$, with $\sigma^\text{MB}=\sqrt{k_B T/m}$, we are more likely to \emph{observe} particles that are moving towards the spacecraft.  We consider only positive velocities and have the probability of observation $p(v|\bm{\theta}) \propto v\, \mathcal{N}(v-u,\sigma^\text{MB});\, v>0$.

There exist closed-form estimators for the quantities $T$ and $u$, which we provide below.  The variance $\Sigma_{i,j}$, with $(i,j)\in\{T,u\}$, of the best possible unbiased estimators is set by the Cram\'er--Rao bound as the reciprocal of the Fisher information matrix $\Sigma=\mathcal{I}^{-1}/N$, where $N$ is the number of measurements, with elements of this matrix computed as  $\mathcal{I}_{i,j}=-\int_0^\infty p(v|\bm{\theta}) ~\partial_{i,j}[\log p(v|\bm{\theta})] \,dv$ \citep{kay1993fundamentals}.

LEVITAS is expected to operate most effectively when the spacecraft is fast compared with the speed of the gas species; we thus consider the limit $u\gg\sigma^\text{MB}$.  In this limit, $\mathcal{I}$ is diagonal, meaning the estimates of $T$ and $u$ are uncorrelated, and we obtain the uncertainties $\text{std}(T)=T\sqrt{2/N}$ and $\text{std}(u)=\sigma^\text{MB}/\sqrt{N}$.

We now seek the maximum likelihood estimators for $T$ and $u$ based on $N$ independent observations of the recoil speed $\{v\}$. Note that in this simplified model, with no measurement noise, we access the speed of the gas atoms directly, and for simplicity of notation we use $v$ rather than $p_\text{I}'$ and related symbols.
In the regime where $u/\sigma^\text{MB}\gg1$, we obtain the simplified normalised probability distribution $p(v|\bm{\theta}) = (v/u)\mathcal{N}(v-u,\sigma^\text{MB})$ for $v\geq0$.

The joint probability for observing the data set $v_i$ is $\prod_{i=1}^N p(v_i|\bm{\theta})$. The log-likelihood is $\mathcal{L}=\sum \log p(v_i|\bm{\theta})$. The estimators are the values which maximise this likelihood, found as the simultaneous solutions to $\partial_T \mathcal{L}=0$ and $\partial_u\mathcal{L}=0$.

These expressions yield, respectively, $k_B T'/m = \langle (v_i-u')^2\rangle$ and $k_B T'/m = u'\langle (v_i-u')\rangle$ where $T',u'$ are the estimated values and $\langle\ldots\rangle$ denotes the sample mean. Equating, expanding out the squared summation, and grouping terms, we find $2u'^2 - 3u' \langle v_i\rangle + \langle v_i^2\rangle=0$. This quadratic has solution
\begin{equation}
u' = \frac{3}{4}\langle v_i\rangle \pm\frac{1}{4}\sqrt{9\langle v_i\rangle^2 - 8\langle v_i^2\rangle}
\end{equation}
and the positive solution is appropriate;
from this the estimator $T'$ follows straightforwardly.
For data satisfying the simplifications of single species and $u/\sigma^\text{MB}\gg1$, these closed-form estimators have been verified to saturate the Cram\'er--Rao bound and agree with the full numerical likelihood maximization described in Sec.~\ref{sec:gas_dynamics}.

\bibliography{references}{}

\end{document}